\documentclass[preprint2]{aastex62}
\usepackage{natbib}
\usepackage{amsmath}
\usepackage{color}
\usepackage[colorlinks=true,
            linkcolor=black,
            urlcolor=black,
            citecolor=blue, breaklinks=true]{}
\usepackage{url}
\usepackage{breakurl}

\shorttitle{Electron Density and Environment at $z=1.62$}
\shortauthors{Harshan et al.}
\newcommand{\mstar}{$\log ({\rm M}_*/{\rm M}_{\odot})$}

\newcommand{\sfr}{$\rm{M}_{\odot}{\rm{yr}}^{-1}$}

\newcommand{\arcse}{$^{\prime\prime}$}

\newcommand{\zfire}{{\tt ZFIRE\ }}

\newcommand{\ukidds}{{\tt UKIDDS}}
\newcommand{\ukirt}{{\tt UKIRT\ }}
\newcommand{\candels}{{\tt CANDELS\ }}
\newcommand{\sdss}{{\tt SDSS}}
\newcommand{\mosdef}{{\tt MOSDEF}}

\newcommand{\oii}{{[\hbox{{\rm O}\kern 0.1em{\sc ii}}]}}
\newcommand{\sii}{{[\hbox{{\rm S}\kern 0.1em{\sc ii}}]}}
\newcommand{\ha}{{\hbox{{\rm H}\kern 0.1em{\sc $\alpha$}}}}
\newcommand{\hb}{{\hbox{{\rm H}\kern 0.1em{\sc $\beta$}}}}
\newcommand{\oiii}{{[\hbox{{\rm O}\kern 0.1em{\sc iii}}]}}
\newcommand{\nii}{{[\hbox{{\rm N}\kern 0.1em{\sc ii}}]}}

\newcommand{\nel}{{$n_{\rm{e}}$}}
\newcommand{\lambdaoii}{{$\lambda3726$,$\lambda3729$}}
\newcommand{\lambdasii}{{$\lambda6717$,$\lambda6731$}}
\newcommand{\wnii}{{$\lambda6584$}}
\newcommand{\woiii}{$\lambda5007$}
\newcommand{\cm}{{cm$^{\rm{-3}}$}}
\newcommand{\kms}{{km\,s$^{\rm{-1}}$}}
\newcommand{\ho}{{km\,s$^{\rm{-1}}$\,Mpc$^{\rm{-1}}$}}
\newcommand{\zspec}{{$z_{\rm{spec}}$}}
\newcommand{\cmmnt}[1]{}
\newcommand{\udscls} {{$366\pm84$}}
\newcommand{\udsfl}{{$104\pm55$}}
\newcommand{\fldmel}{{$113\pm63$}}
\newcommand{\udsall}{{$254\pm76$}}
\graphicspath{{./}{figures/}}

\begin{document}
\title{{ \tt ZFIRE}: Measuring Electron Density with \oii\ as a function of environment at $z=1.62$}
\author[0000-0001-9414-6382]{Anishya Harshan}
\affiliation{School of Physics, University of New South Wales, Sydney, NSW 2052, Australia}
\affiliation{ARC Centre of Excellence for All Sky Astrophysics in 3 Dimensions (ASTRO 3D), Australia}

\author[0000-0002-8984-3666]{Anshu Gupta}
\affiliation{School of Physics, University of New South Wales, Sydney, NSW 2052, Australia}
\affiliation{ARC Centre of Excellence for All Sky Astrophysics in 3 Dimensions (ASTRO 3D), Australia}

\author[0000-0001-9208-2143]{Kim-Vy Tran}
\affiliation{School of Physics, University of New South Wales, Sydney, NSW 2052, Australia}
\affiliation{ARC Centre of Excellence for All Sky Astrophysics in 3 Dimensions (ASTRO 3D), Australia}

\author[0000-0002-2250-8687]{Leo Y. Alcorn}
\affil{Department of Physics and Astronomy, Texas A\&M University, College Station, TX, 77843-4242 USA}
\affil{George P.\ and Cynthia Woods Mitchell Institute for Fundamental Physics and Astronomy, Texas A\&M University, College  Station, TX, 77843-4242}
\affiliation{Department of Physics and Astronomy, York University, 4700 Keele Street, Toronto, Ontario, ON MJ3 1P3, Canada}

\author[0000-0002-9211-3277]{Tiantian Yuan}
\affiliation{Swinburne University of Technology, Hawthorn, VIC 3122, Australia}
\affiliation{ARC Centre of Excellence for All Sky Astrophysics in 3 Dimensions (ASTRO 3D), Australia}

\author[0000-0003-1362-9302]{Glenn G. Kacprzak}
\affiliation{Swinburne University of Technology, Hawthorn, VIC 3122, Australia}
\affiliation{ARC Centre of Excellence for All Sky Astrophysics in 3 Dimensions (ASTRO 3D), Australia}

\author[0000-0003-2804-0648]{Themiya Nanayakkara}
\affiliation{Leiden Observatory, Leiden University, P.O. Box 9513, NL 2300 RA Leiden, The Netherlands}

\author[0000-0002-3254-9044]{Karl Glazebrook}
\affiliation{Swinburne University of Technology, Hawthorn, VIC 3122, Australia}

\author[0000-0001-8152-3943]{Lisa J. Kewley}
\affiliation{Research School of Astronomy and Astrophysics, The Australian National University, Cotter Road, Weston Creek,
	ACT 2611, Australia}

\affiliation{ARC Centre of Excellence for All Sky Astrophysics in 3 Dimensions (ASTRO 3D), Australia}
\author[0000-0002-2057-5376]{Ivo Labb\'e}
\affiliation{Swinburne University of Technology, Hawthorn, VIC 3122, Australia}

\author[0000-0001-7503-8482]{Casey Papovich}
\affiliation{Department of Physics and Astronomy, Texas A\&M University, College Station, TX, 77843-4242 USA}
\affiliation{George P.\ and Cynthia Woods Mitchell Institute for Fundamental Physics and Astronomy, Texas A\&M University, College  Station, TX, 77843-4242}

\begin{abstract}
The global star formation rates (SFR) of galaxies at fixed stellar masses increase with redshift and are known to vary with environment up to $z\sim2$. We explore here whether the changes in the star formation rates also apply to the electron densities of the inter-stellar medium (ISM) by measuring the \oii\ (\lambdaoii) ratio for cluster and field galaxies at $z\sim2$. We measure a median electron density of  \nel\,=\,\udscls\,\cm\ for six galaxies (with $1\sigma $ scatter = 163\,\cm) in the UDS proto-cluster at $z=1.62$. We find that the median electron density of galaxies in the UDS proto-cluster environment is three times higher compared to the median electron density of field galaxies (\nel\,=\,\fldmel\,\cm\ and $1\sigma $ scatter = 79\,\cm) at comparable redshifts, stellar mass and SFR. However, we note that a sample of six proto-cluster galaxies is insufficient to reliably measure the electron density in the average proto-cluster environment at $z\sim2$. We conclude that the electron density increases with redshift in both cluster and field environments up to $z\sim2$ (\nel\,=\, $30\pm1$\,\cm\ for $z\sim0$ to \nel\,=\,\udsall\,\cm\ for $z\sim1.5$). We find tentative evidence ($\sim 2.6\sigma$) for a possible dependence of electron density on environment, but the results require confirmation with larger sample sizes.

\end{abstract}

\section{Introduction}
Environment plays an extensive role in the evolution of galaxies. In the low-redshift universe ($z<0.2$), high-density or cluster environment show a higher fraction of quenched galaxies and have galaxies with lower gas fractions compared to low-density or field environment \citep{Iraoka2000, Nichol2003, Kauffmann2004,  Blanton2006, Lewis2008, chung2009, Ellison2009, Barsanti2018, Grootes2018,Koyama2018,Davies2019}. The frequency of lenticular and elliptical galaxies increases, and the frequency of spiral galaxies decreases with the local density indicating that environment affects the morphology of galaxies \citep{Dressler1980, VanDerWel2009, Sobral2011, Houghton2015,Paulino-Afonso2019a}.

One possible explanation for the observed differences is that in high-density environments, the probability of galaxy-galaxy interactions (collisional and tidal interactions) increases. Through galaxy-galaxy interactions and interactions with the intra-cluster medium (ICM), star-forming disk galaxies transform into quenched spheroidals \citep{Gunn1972, Moore1996, Gnedin2003, Smith2005}.   

As galaxies fall into the cluster, gas is stripped off through ram pressure stripping \citep{Gunn1972, Balogh2004, Hester2006, Cortese2009, Nichols2011, Brown2017, Gupta2017}, resulting in a gradual decline in the SFR as galaxies run out of their star formation fuel \citep[strangulation;][]{Peng2015, Bahe2015, Wang2018}. Both simulations and observational studies find evidence of lower star formation in cluster galaxies compared to field galaxies up to $z\sim2$ \citep{lewis2002, McGee2011,Rasmussen2012, Tran2015, Paccagnella2016, Bahe2017, Genel2018, Sobral2016,Darvish2016, Darvish2017,Muzzin2018, Davies2019,Paulino-Afonso2019a}. At redshift $z=1.62$, \cite{Tran2015} find systematically lower star formation rates in the UDS (Ultra-Deep Survey) proto-cluster galaxies compared to the field galaxies, indicating a tentative effect of environment albeit not statistically significant.

Existing studies show that star-forming galaxies at redshift $z>1$ have higher electron densities \citep{Brinchmann2008, Bian2010, Shirazi2013} than their local counterparts. Electron densities of star-forming galaxies (SFGs) at $z>1$ show significant correlation to global galaxy properties such as SFR and specific SFR (sSFR) \citep{Kaasinen2016, Shimakawa2015} but no significant correlation with the ionization parameter \citep{Shimakawa2015}. Because electron density of a galaxy varies with the SFR and sSFR \citep{Shimakawa2015, Kashino2017}, variation of electron density with environment needs to be further explored.

%\textit{Because electron density of a galaxy varies with the SFR and sSFR \citep{Shimakawa2015, Kashino2017}, we hypothesise that the electron density also varies with environment. The electron density measurements have been limited in galaxy clusters at $z<0.2$, where the fraction of star-forming galaxies with emission lines is less than $10\%$ \citep{lewis2002, Davies2019}.} 

The electron density measurements have been limited in galaxy clusters at $z<0.2$, where the fraction of star-forming galaxies with emission lines is less than $10\%$ \citep{lewis2002, Davies2019}. At $z\sim$0.5, there are indications that the electron density depends on the local environment \citep{Sobral2015,Darvish2015}.

\cite{Darvish2015} find a negative correlation between the electron density of galaxies and their local environment density at $z\sim0.5$. They find that electron density of low  stellar mass galaxies in the filamentary structure is nearly 17 times lower than the electron density of field galaxies at the same stellar mass, SFR and sSFR. 

Whereas at redshift $z>1$, low signal-to-noise and insufficient sample size limits electron density measurements as a function of environment. With the advent of sensitive near-infrared and optical spectrographs, we can now probe the ``redshift desert" \citep[$1<z<3$][]{Steidel2014,Kacprzak2015,Nanayakkara2016,Harrison2017,Turner2017}. Extensive studies are done on effects on environment on mass-metallicity relation, BPT diagnostics and star formation, however environmental effects on electron density studies still remain largely unexplored at higher redshifts  \citep[$z>1$; ][]{1981PASP...93....5B, Tran2003, Bassett2013, Sobral2013, Kewley2015, Wuyts2016, Turner2017, Alcorn2019} .

In our paper, we investigate the effect of environment on the electron density in the UDS proto-cluster at redshift $z=1.62$ \citep[confirmed by][]{Papovich2010,Tanaka2010, Tran2015}. We use Keck-LRIS observations of the UDS proto-cluster taken as part of the \zfire survey \citep{Tran2015,Nanayakkara2016}. We estimate electron density using the \oii \ (\lambdaoii) emission line doublet observations of the UDS proto-cluster.

Our paper is organized as follows. In section \ref{sec:Data}, we describe the selected sample and data reduction process. We describe the method of electron density estimation in section \ref{subsec:elden} and state our results and analysis in section \ref{sec:results}. We discuss and summarize our results in section \ref{sec:disc} and \ref{sec:sum}.

For this work, we assume a flat $\Lambda$CDM cosmology with $\Omega_{M}$\,=\ 0.3, $\Omega_{\Lambda}$\,=\ 0.7, and H$_{0}$\,=\ 70\,\ho . At redshift $z = 1.62$, $1$\arcse \, corresponds to an angular scale of 8.47 kpc.

\section{DATA and Methodology}
\label{sec:Data}
\subsection{UDS Cluster}
\label{subsec:UDS cluster}

Our sample is sourced from the \zfire survey \citep{Tran2015, Nanayakkara2016}, which combines optical and near infrared spectroscopy of the proto-cluster in the UDS field at redshift $z_{cl}=1.623$. The spectroscopic targets for the \zfire survey were selected from the UDS catalog \citep{Williams2009} created as a part of the \ukirt Infrared Deep Sky Survey (\ukidds), a near infrared imaging survey \citep{Lawrence2007}\footnote{ UDS proto-cluster also referred as XMM-LSS J02182-05102 or IRC 0218 \citep{Tran2015} and CLG0218.3-0510 \citep{Tran2010} and \citep{Santos2014} }.

The UDS proto-cluster, first reported by \cite{Papovich2010,Tanaka2010} is one of the first cluster used to demonstrate an increase in star formation density with local galaxy density \citep{Tran2010}. Still in its formative phase  \citep{Rudnick2012}, the UDS proto-cluster has total star formation rate $>1000$\,\sfr\ \citep{Santos2014} and is an ideal candidate to study the variation of galaxy properties in high-density environments at $z>1.5$.

%\textit{The UDS proto-cluster has total star formation rate $>1000$\,\sfr\ \citep{Santos2014} and is still in its formative phase \citep{Rudnick2012} and is an ideal candidate to study the variation of galaxy properties in high-density environments at $z>1.5$.}

Using the Keck-LRIS and Keck-MOSFIRE spectroscopy, 33 cluster members are identified in the redshift range $1.6118 \leq z_{\rm{spec}} \leq 1.6348$. The median redshift of the proto-cluster is $z_{\rm{cl}} = 1.623 \pm 0.0003$ and the cluster velocity dispersion is $\sigma_{\rm cl} = 254 \pm 50 $ \kms\  \citep{Tran2015}.

\subsection{Optical Spectroscopy: Keck-LRIS }
\label{subsec:spec}
The optical observations were carried out as a part of the \zfire survey on the Low Resolution Imaging Spectrometer  \citep[LRIS;][]{Oke1995} with a $5.5'\times8'$ field of view and resolution of $0.135$\arcse \, per pixel. LRIS is equipped with red and blue cameras that can simultaneously cover a wavelength range of $3200$\,\AA$\, - \,10000$\,\AA. The primary targets were candidate star-forming cluster galaxies identified by \cite{Tran2015}, candidate Lyman-Break Galaxies at $z_{\rm{phot}} $\,$>1.35$, and \oii\ emitters identified by \cite{Tadaki2012} from narrow-band imaging with magnitude $i_{AB}<21 $ mag. The secondary targets and mask fillers were galaxies with magnitude $21<i_{AB}<24$. 

Observations were taken in excellent conditions with median seeing of about $0.6$\arcse\ on 19 and 20 October 2012 (NASA/Keck Program ID 48/2012B). Brightest cluster galaxies were targeted with high priority and observed in 3 out of 4 masks with $9 \times 20 $ minute exposures. The fourth mask with low priority targets was observed for $5 \times 20$ minute exposures. In 4 masks, we observed a total of 136 galaxies.

The blue side of the spectrum covers a wavelength range $3800 $\,\AA$<\lambda<5800$\,\AA\ using 600/4000 grism, and the red side $7000 $\,\AA\,$ < \lambda <10000 $\,\AA\ using 600/10000 grating. A slit width of $1$\arcse \, results in a spectral resolution of $4.0 $\ \AA\   and $4.7$\ \AA\ for the blue and red spectra respectively. With a resolution of $4.7 $\,\AA\, in the observed frame, we get resolution of $1.79 $\,\AA\, at $z = 1.62$ rest-frame. The \oii (\lambdaoii) doublet at $z \approx 1.62$ is observed at wavelength range approximately $9760 $\,\AA\,to $9770 $\,\AA\,and the $2.7 $\,\AA\,rest-frame wavelength separation should be resolved with the $1.79 $\,\AA\,resolution.

Spectra were reduced using IRAF routines with custom software provided by D. Kelson \citep{Kelson2003} for the red and blue sides separately. Cosmic ray rejection on the red side was done using \textit{crutil} in IRAF. Median rectified science images after flat-fielding, wavelength calibration and sky line correction were used to create the combined images \citep{Tran2015}.

\subsection{1-D Spectral Extraction }
\label{subsec:1d}
We extract 1-D spectra from the reduced red side of the 2D spectrum from LRIS-Keck by summing over the entire slit length and de-redshifting it to rest-frame using the photometric redshift taken from \cite{Tran2015}. On the extracted initial 1-D spectrum, we fit a double Gaussian profile using the \textit{optimize.curvefit} routine from the scipy library in Python to calculate spectroscopic redshift (\zspec). We de-redshift the spectrum in the initial step to provide a reliable set of first-guesses for the double gaussian parameters to the fitting routine \textit{optimize.curvefit}.

To identify the peak in the spatial direction, we select the wavelength window such that $3\sigma$ of the flux from \oii\ doublet is included. We collapsed the spectrum in the selected wavelength window along the spatial direction and fit a Gaussian profile to the extracted spatial profile. This is done to reduce the contamination by the sky absorption lines very close to the \oii\ emission lines. We take $3\sigma$ region around the centroid of the best-fit Gaussian profile as the position of galaxy along the slit and collapse the 2-D spectra in the selected spatial region (shown by purple lines in  Fig.\,\ref{fig:cluster} and Fig.\,\ref{fig:field}) along the wavelength direction to extract the 1-D spectrum for each galaxy. We visually inspect all apertures to ensure the inclusion of the both emission lines.

To minimize the effect of rotation and to remove spectral regions in the galaxy with blended \oii\ lines, we modify the window in which we collapse the 2-D spectra for several galaxies. Purple lines in Fig.\, \ref{fig:cluster} (cluster galaxies) and Fig.\,\ref{fig:field} (field galaxies) show the window selected where 2-D spectra is collapsed to extract the 1-D spectra. We select a smaller aperture to avoid the regions of blended emission lines. In the region with blended \oii\ lines, we cannot extract along rotational axis because it would introduce further uncertainties. Selecting small aperture will not affect the calculation of electron density as the doublet lines are visually congruent and thus the ratio of two emission lines would remain constant.

\subsection{Emission Line Fitting}
\label{subsec:linefit}
We use reduced red side of the 2-D spectrum comprising of wavelength range $7000 $\,\AA\,to $10000 $\,\AA\, of the Keck-LRIS data of the Ultra-Deep Survey (UDS) field using the method  defined in \citet{Tran2015}. We also use the redshift catalogs created by \cite{Tran2015}.

%\textit{With a resolution of $4.7 $\,\AA\, in the observed frame, we get resolution of $1.79 $\,\AA\, at $z = 1.62$ rest-frame. The \oii (\lambdaoii) doublet at $z \approx 1.62$ is observed at wavelength range approximately $9760 $\,\AA\,to $9770 $\,\AA\,and the $2.7 $\,\AA\,rest-frame wavelength separation should be resolved with the $1.79 $\,\AA\,resolution. We also use the redshift catalogs created by \cite{Tran2015}.}

While fitting the double Gaussian profile, we constrain the separation of the two peaks to be $2.7 $\,\AA\,in the rest-frame as measured by atomic physics and require line widths of the two lines to be the same. We tested the fitting by relaxing the constraint on the separation between the \oii\ emission lines by $0.5 $\,\AA\,  but found no significant difference in the flux ratios. We weight the fit with the sky residual spectrum to reduce the effects of sky absorption. We measure the flux by integrating the fitted Gaussian profile within $3\sigma$ bound for each emission line. To determine the uncertainty in the electron density, we generate 500 Gaussian random spectra by perturbing the flux at each wavelength according to the sky noise at that wavelength. We calculate the \oii\ doublet fluxes for each generated spectra and take the standard deviation of the created fluxes to be the $1\sigma$ error for each emission line flux.

\subsection{Galaxy Selection}
\label{subsec:galsel}
Due to the presence of many sky absorption lines in the rest-frame wavelength window near the \oii\ emission lines, we select a sub-sample of galaxies by visually assigning each galaxy a quality flag  $Q : 0-3$ that indicates the quality of the observation. Galaxies with barely visible emission lines or where lines are contaminated with sky absorption are rated 0. Galaxies with quality rating of 3 are the ones with clearly resolved doublet emission and minimal rotation in the selected aperture as shown in Fig.\,\ref{fig:cluster} and Fig.\,\ref{fig:field}. For our study, we only consider the galaxies with Q = 3 rating, which results in a sample of 8 galaxies in the redshift regime of $1.3 \leq z \leq 1.7 \ (1\,Gyr)$. Out of the 8 galaxies, 6 are proto-cluster member galaxies because they lie in the redshift range $1.6118 \leq z \leq 1.6348$ \citep{Tran2015} and rest are field galaxies. 

\begin{figure}[h] 
\centering
\includegraphics[scale=0.35]{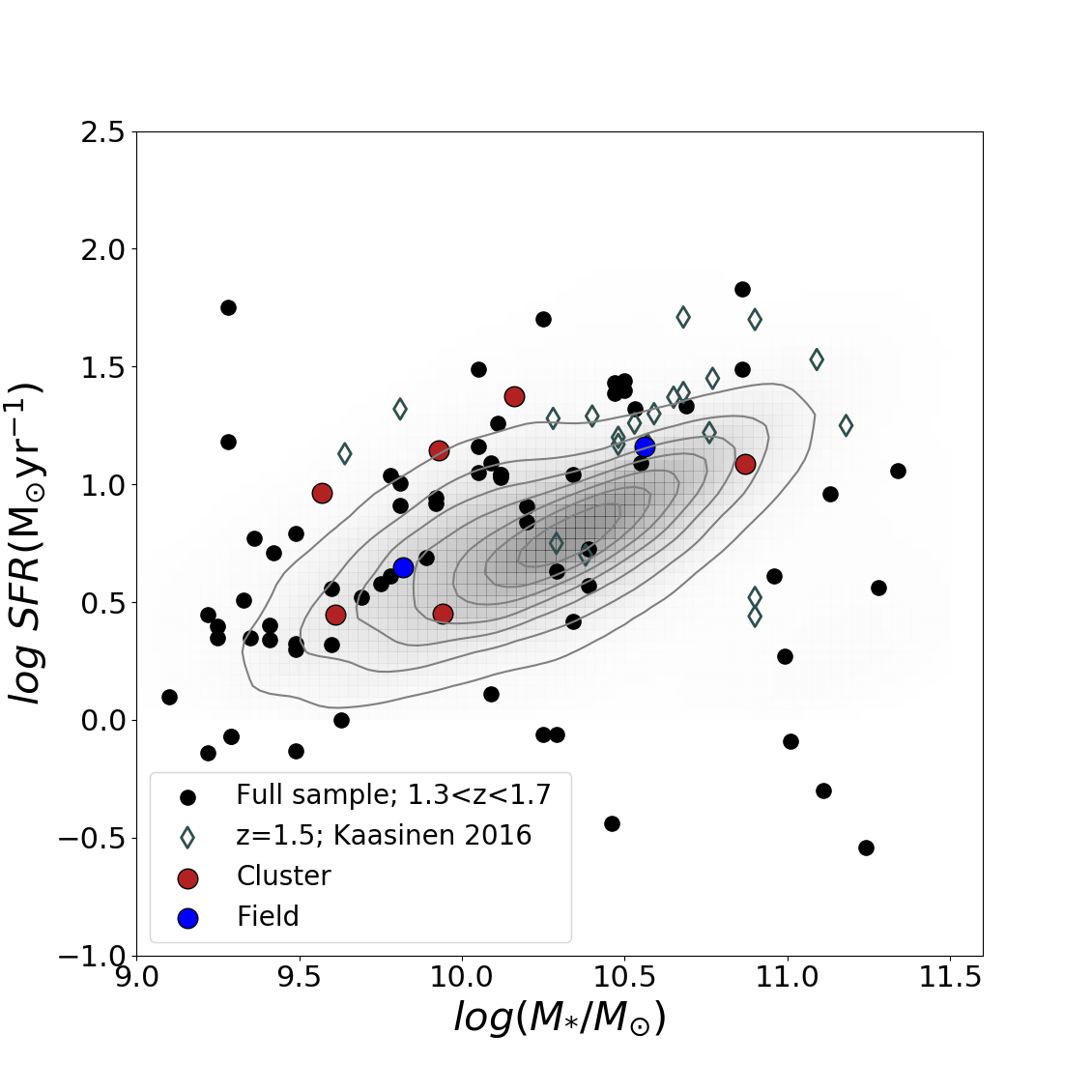}
\caption{SFR vs Stellar Mass for the full sample at $1.3<z<1.7$ (black dots), selected cluster galaxies (red circles), selected field galaxies (blue circles), comparison sample at $z\sim 1.5$ from \cite{Kaasinen2016}(open diamonds) and the full SDSS sample (grey meshed contours). For high redshift samples, electron density is calculated with \oii\ and for the local sample (SDSS), electron density is measured using \sii. } 
\label{fig:sfr_mass}
\end{figure}

Fig.\,\ref{fig:sfr_mass} shows the SFR - stellar mass relation for the full sample, selected sub-sample with a quality rating of three and the comparison samples.  Due to  observational limitations and selection effects, all high redshift galaxies in the sample are biased towards galaxies with higher SFR. The high redshift sample spans the full range in SFR to the local SDSS sample. 
%Selection of galaxies with bright \oii\ emission intrinsically selects galaxies towards the high end of SFR distribution. 
A student's $t-$ test confirms the SFR and stellar mass distribution of the selected sample is consistent with the parent sample with p-values of 0.9 and 0.65. 
%However, both cluster and field samples are selected in the same manner and have the same intrinsic bias.
The SFR and stellar mass distribution of our selected cluster and field samples are also consistent with each other with a p-value of 0.9 and 0.7 respectively. 

\subsection{Local Comparison Data}
\label{subsec:localdata}
Our Local comparison data has been taken from the Sloan Digital Sky Survey (\sdss) - DR7. Stellar masses, star formation rates and specific star formation rates have also been taken from the Galspec data of \sdss\ DR-7 \citep{York2000,Abazajian2009} provided by MPA-JHU group. As the spectra is observed with  $3$\arcse \, aperture and thus do not represent the entire galaxy, the total stellar mass are estimated using ugriz galaxy photometry \citep{Kauffmann2003,Brinchmann2004,Tremonti2004}. To minimize the aperture effects we select galaxies in $0.04 < z < 0.1$ \citep{Kewley2005}. We also reject AGNs from the sample following the \cite{Kauffmann2003a} criteria using optical line ratios \oiii/\hb\ and \nii/\ha. Our Final sample includes 117000 galaxies in the local sample.

We select objects with signal-to-noise ratio $SNR > 3$ for emission lines \oiii\,(\woiii), \hb, \nii\,(\wnii), \ha, \sii\,(\lambdasii). Because the \oii\ doublet is not resolved in the \sdss\ DR7, we calculate electron density with resolved \sii(\lambdasii) doublet. We note that calculating electron density using \sii\ and \oii\ probes different parts of the HII regions of the galaxy \citep{Kewley2019}. However, \cite{Sanders2016} show that electron density calculated with \sii\ is comparable within the uncertainties in our data to the electron density calculated using \oii\ doublet.

  %\textcolor{red}{ \cite{Sanders2016} find that electron density measurements using \oii\ doublet are systematically higher than the \sii\ electron densities. However, the difference is smaller than the measurements uncertainties in our data. } } 

To compare the \sdss\ local galaxy sample with the high redshift sample, we convert the total stellar masses of the low redshift sample from \cite{Kroupa2001}  to \cite{Chabrier2003} initial mass function (IMF) with a constant scaling of 1.06 \citep{Zahid2012}.

\subsection{Comparison Data at $1.5<z<2.6$}
\label{subsec:comdata}
For comparison with redshift $z>1$ we have collected three different data sets. $z\sim 1.5$  sample taken from \cite{Kaasinen2016} consists of galaxies from the COSMOS field between $1.4<z<1.7$. These galaxies are selected to be \oii\ emitters and were observed as part of the COSMOS \oii\ survey. The spectroscopic data has been taken on DEep Imaging Multi-Object Spectrograph (DEIMOS) on Keck II. We select 21 galaxies from the sample that was selected to be \mstar$>9.8$, SFR$_{\rm phot}$ $ \geq 10$\,\sfr\,  and $z(AB)$ magnitude $\lesssim 24 $ . The stellar mass has been converted to Chabrier IMF from Kroupa IMF for comparison with the other cluster sample and the specific Star Formation Rate (sSFR) has been calculated as SFR/Stellar mass for each galaxy. \cite{Yuan2014} find that the structural over-densities in the COSMOS field is at $z = 2.09578 \pm 0.00578$. The \cite{Kaasinen2016} comparison data is outside of the redshift of number over-density in the COSMOS field, so we consider these as field galaxies. 

The  redshift $z=2.3$ sample has been taken from MOSFIRE Deep Evolution Field survey (\mosdef) Survey \citep{Sanders2016}. We take the \oii\,(\lambdaoii) doublet line ratio, stellar mass and SFR from \cite{Sanders2016}. These observations were taken with MOSFIRE on GOODS-S and UDS-CANDELS field. The known over-densities in the UDS-CANDELS field is at $z=1.62$ \citep{Papovich2010, Tanaka2010}  and in GOODS-S is at $z=3.5$ \citep{Forrest2017}. Hence, it is a reasonable assumption that $z\sim2.3$ galaxies in these fields are field galaxies.

Our redshift $z\sim2.5$ sample is taken from the plots in  \cite{Shimakawa2015}. We take electron densities calculated for each \ha\ emitter using the \oii\ doublet emission line ratio and TEMDEN code distributed in the stsdas package and get a sample of 14 galaxies.

\begin{figure*}[t] 
\begin{center}
\includegraphics[scale=0.15]{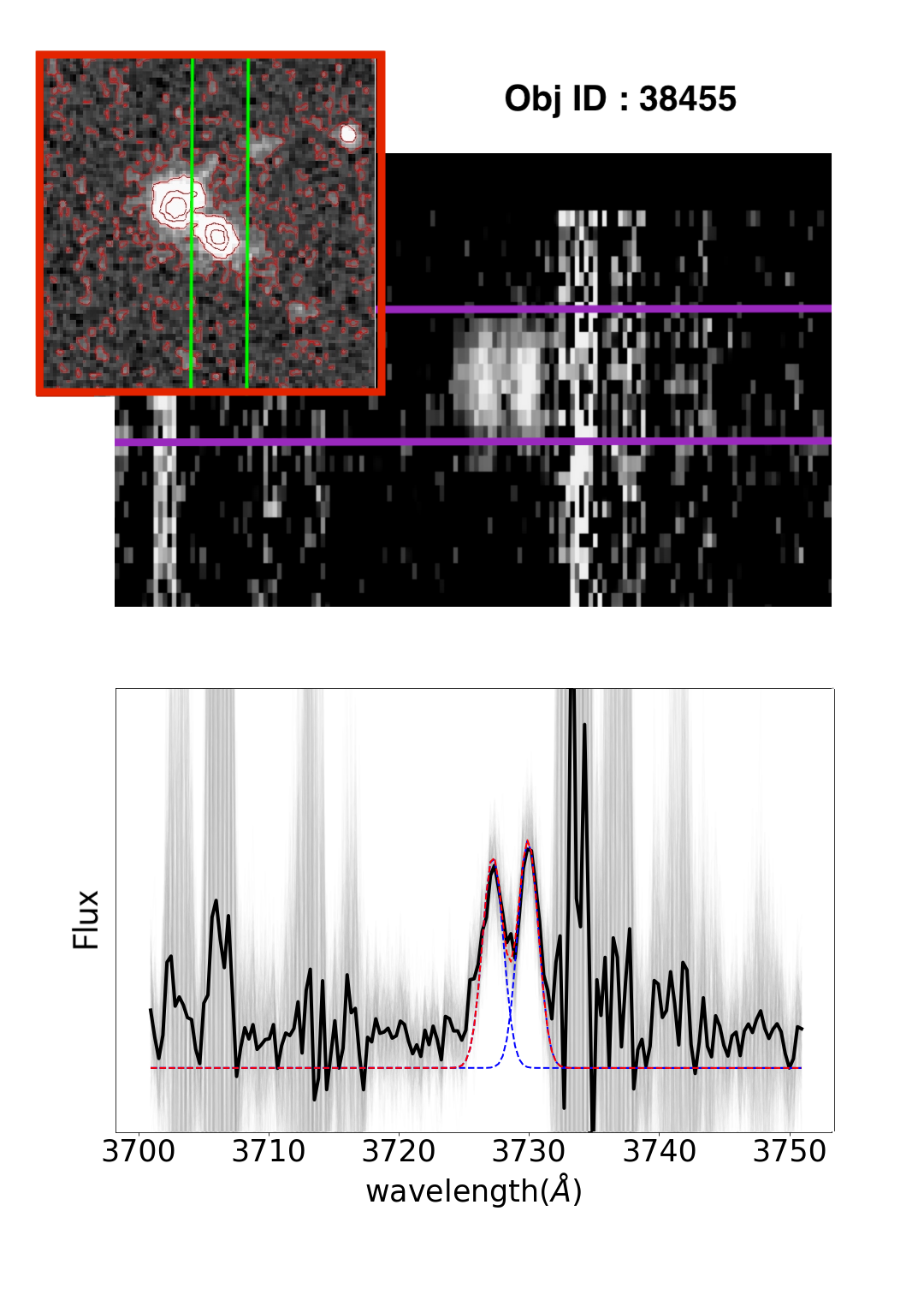}
\includegraphics[scale=0.15]{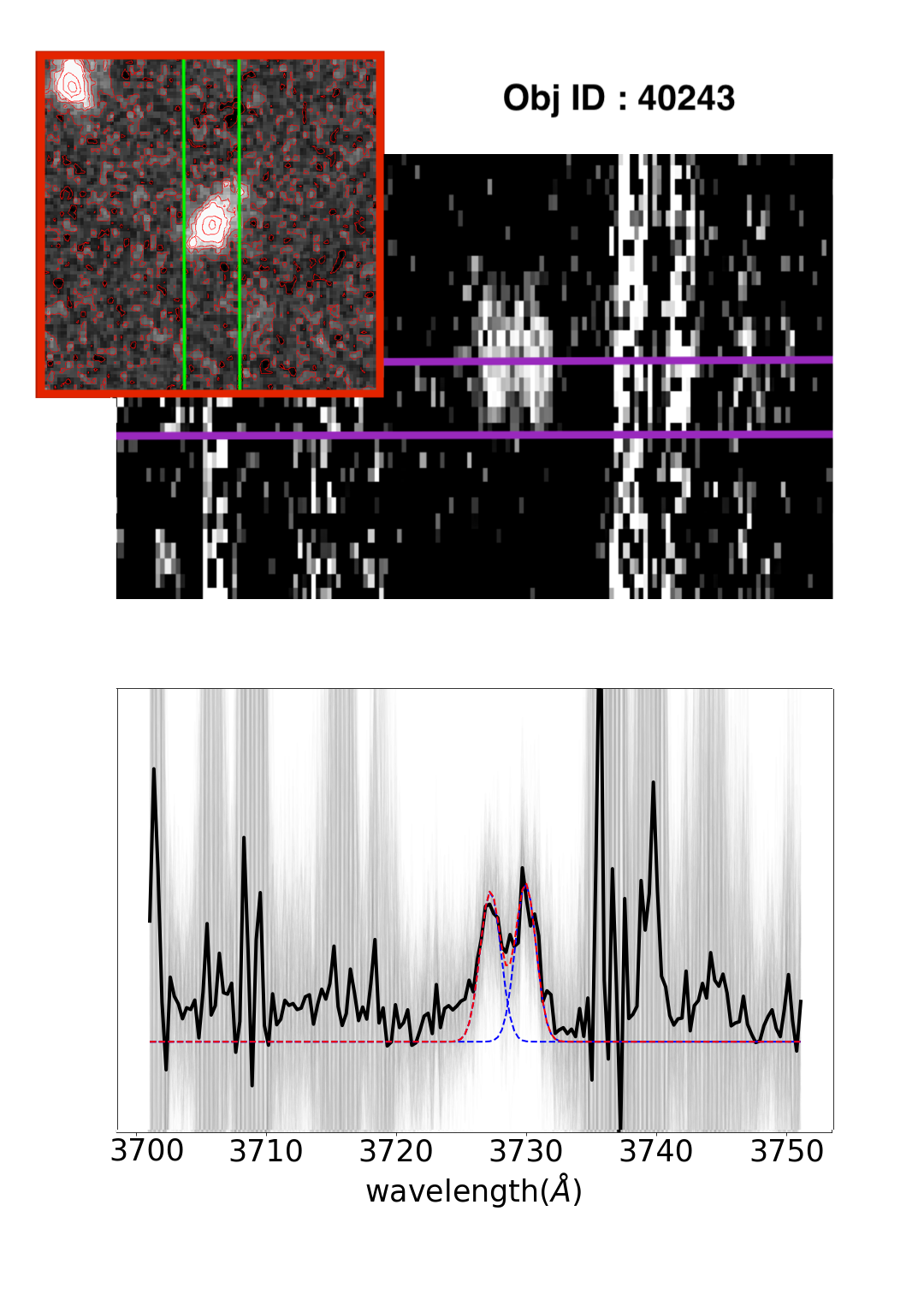}
\includegraphics[scale=0.15]{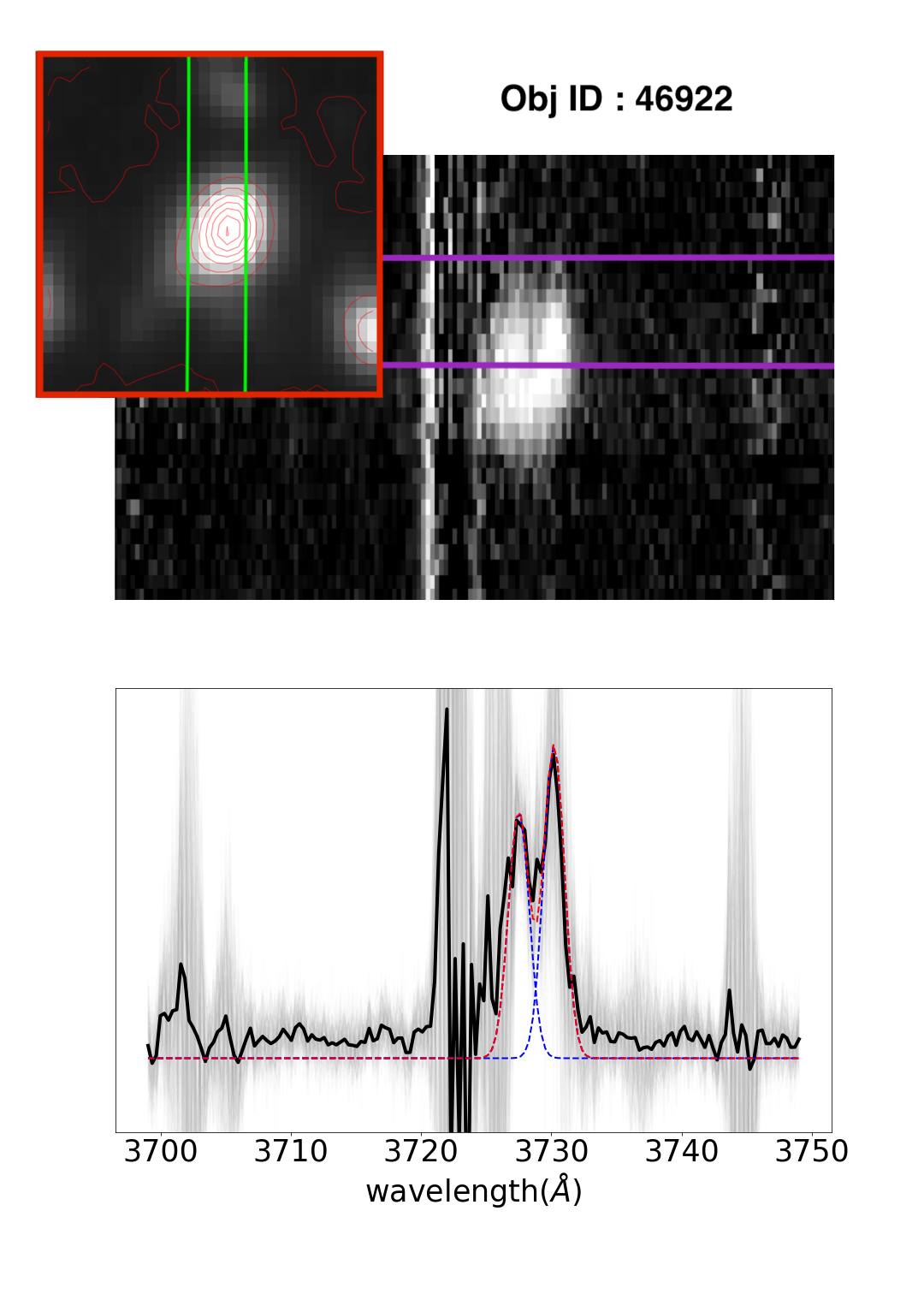}\\
\includegraphics[scale=0.15]{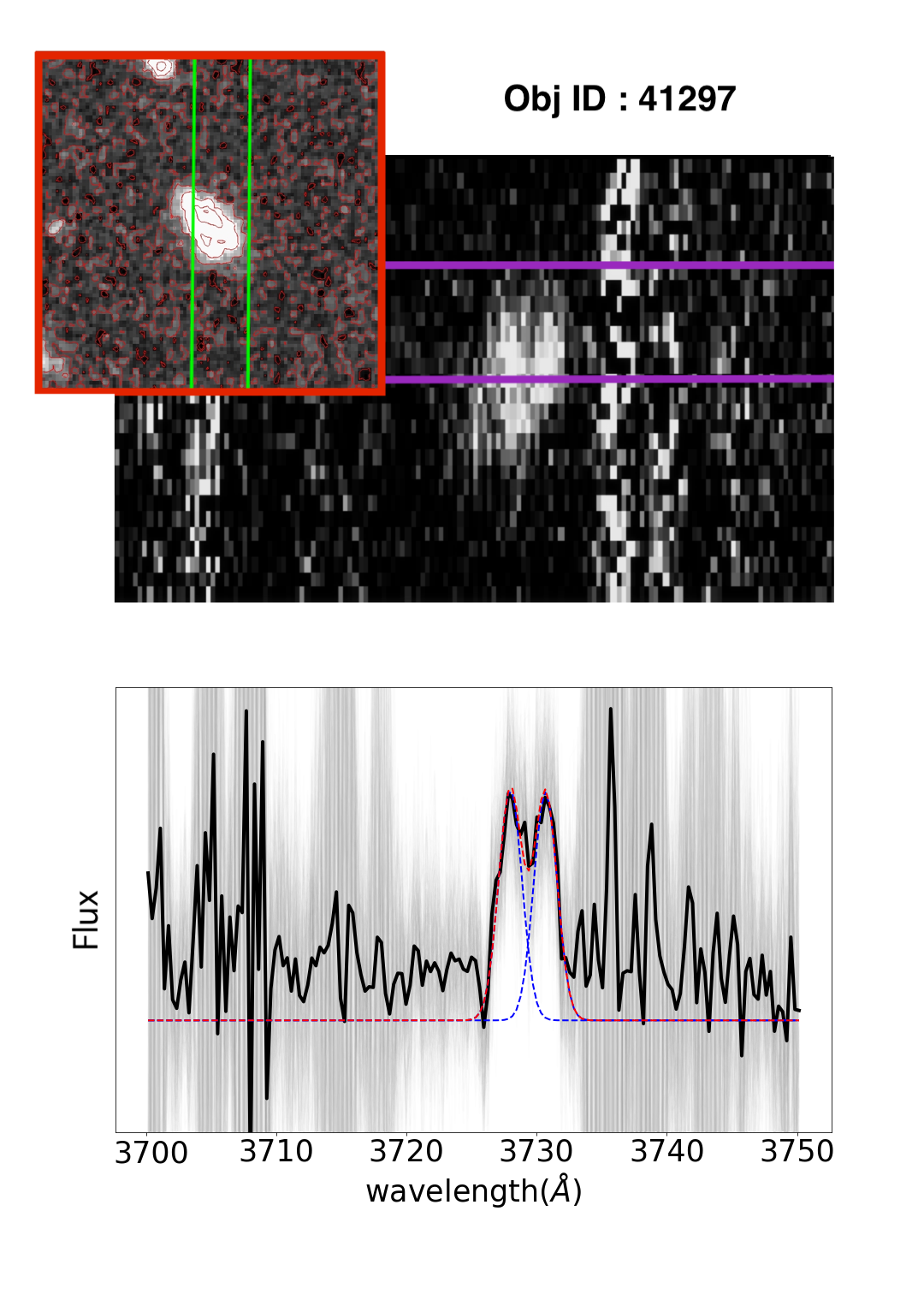}
\includegraphics[scale=0.15]{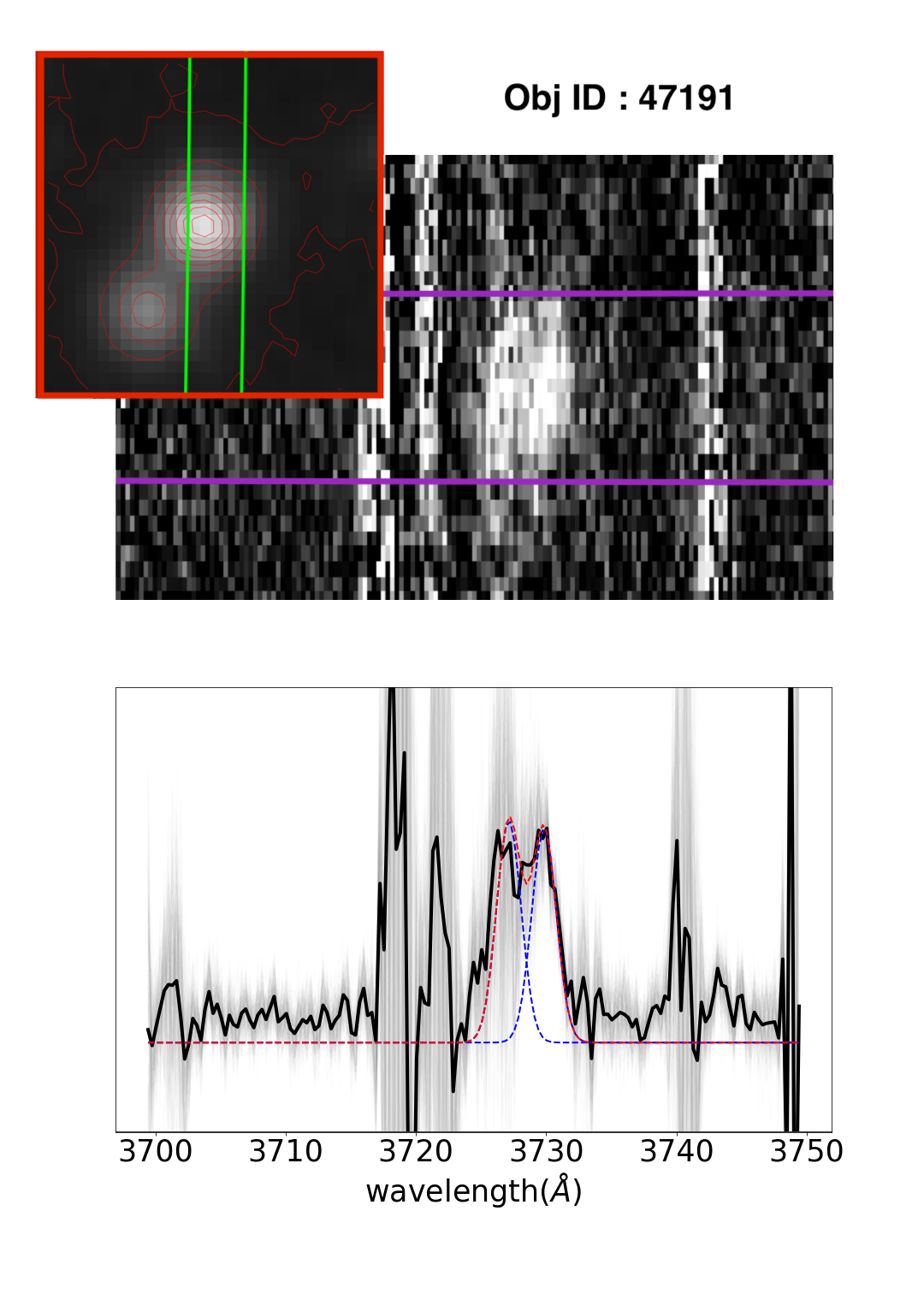}
\includegraphics[scale=0.15]{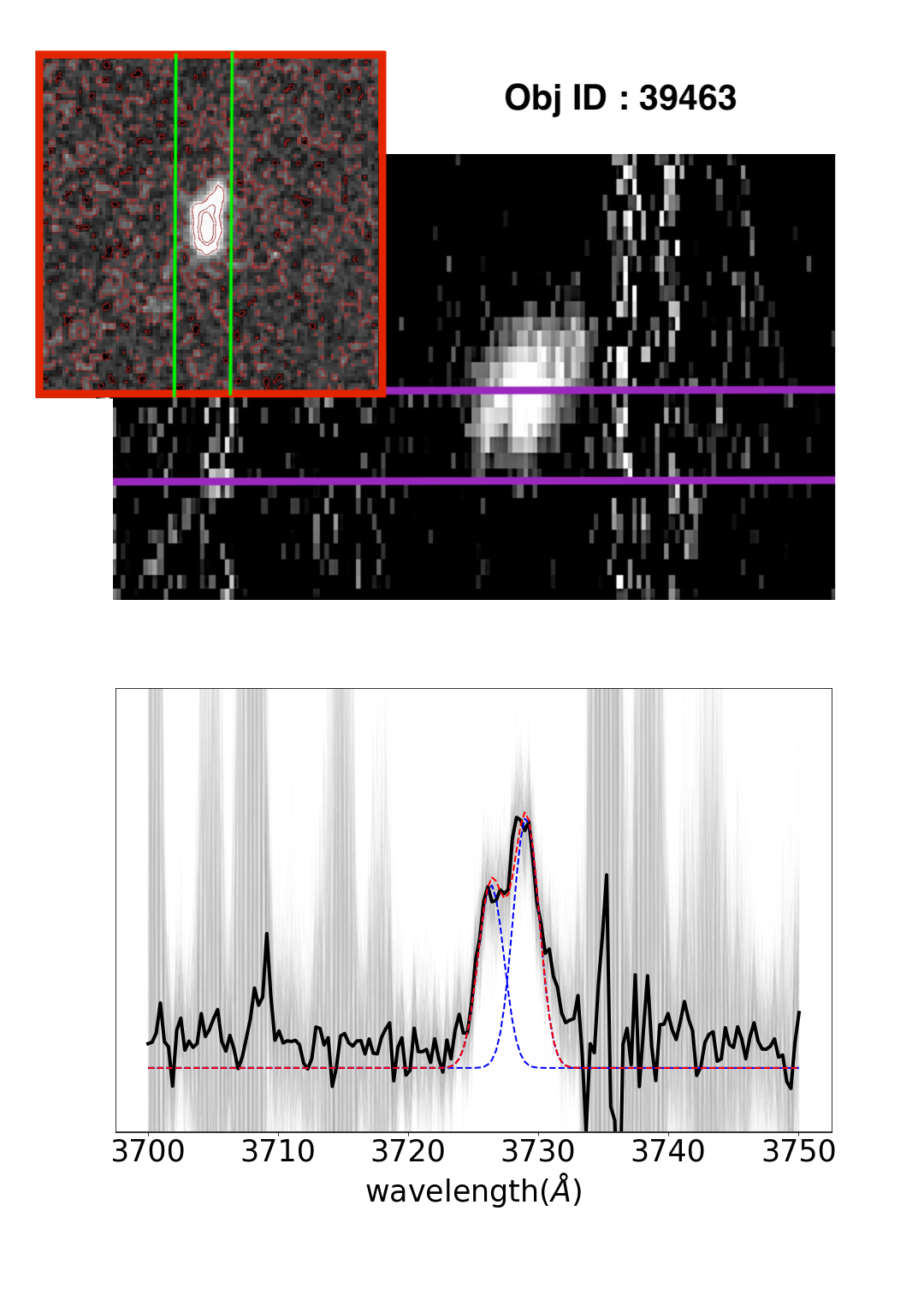}
\caption{Restframe spectra of cluster members within $1.6118 \leq z \leq 1.6348$. The top panel shows the 2D spectrum overlaid with the 6\arcse $\times $ 6\arcse\, HST(F125)/Subaru images. The purple lines show the window of spectra used to extract 1-D spectra. The green lines on the HST(F125)/Subaru(stacked v,b and i band imaages) images are the LRIS slits on the galaxy. The lower panel shows the extracted 1D spectrum inside the aperture shown with purple lines. The grey region show bootstrapped spectra and the black solid line is the median spectrum of the bootstrapped sample. The red dashed line is the fitted double Gaussian profile and the blue dashed line is the fitted Gaussian to each emission line.}
\label{fig:cluster}
\end{center}
\end{figure*}

\begin{figure*}[t] 
\begin{center}

\includegraphics[scale=0.15]{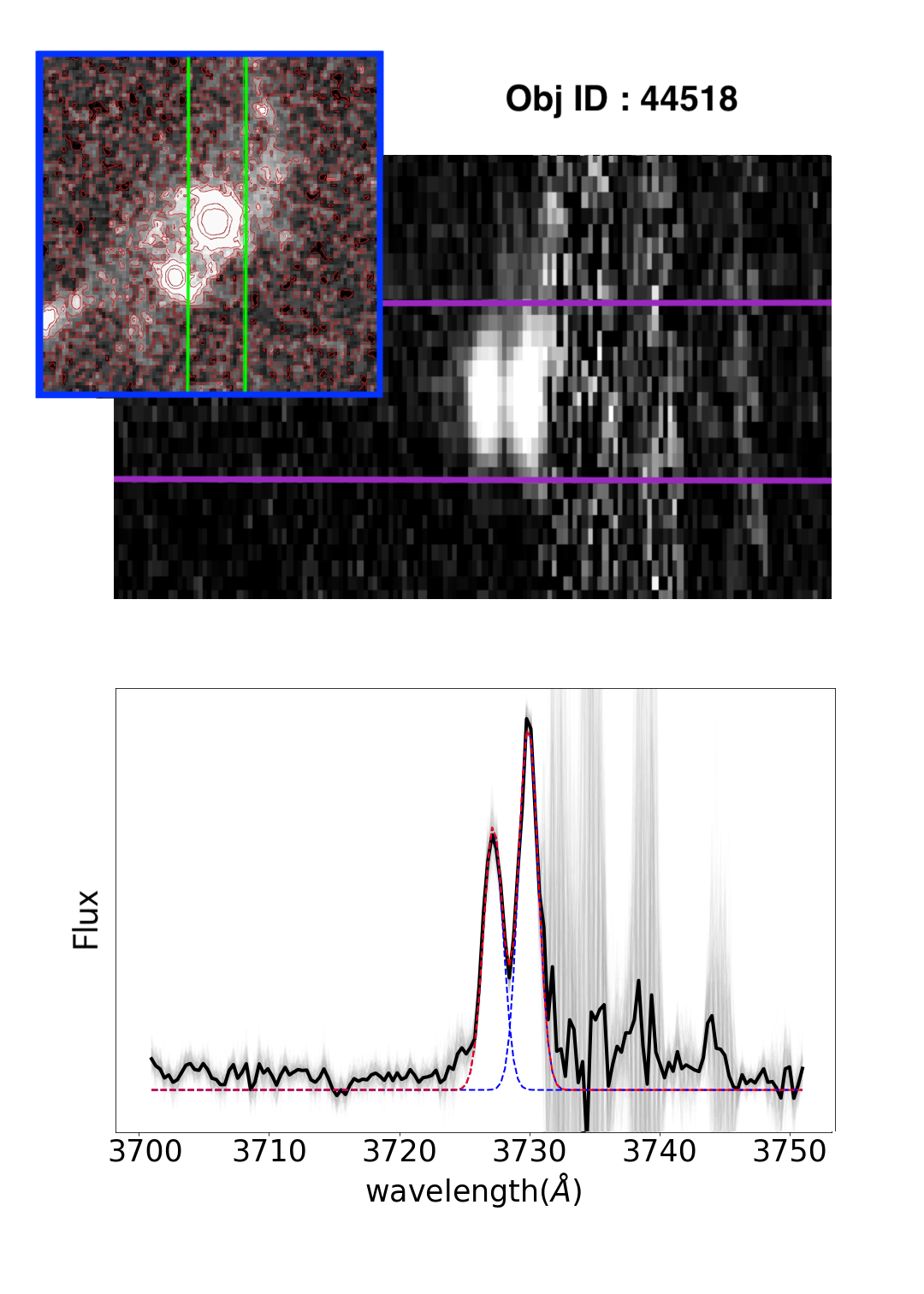}
\includegraphics[scale=0.15]{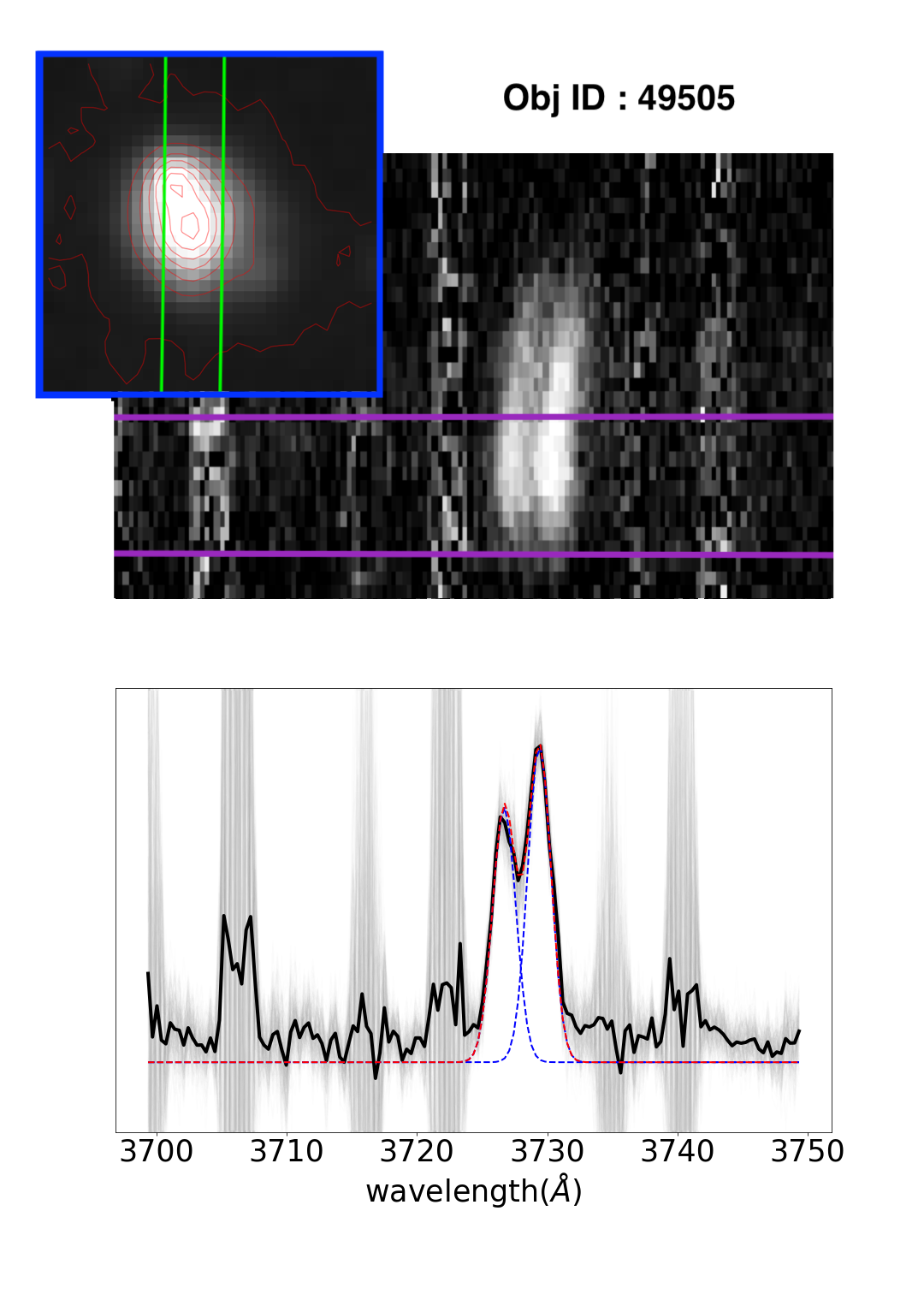}
\caption{Restframe spectrum of field galaxies  within $z \leq 1.6118$ and $z \geq 1.6348$. The top panel shows the 2D spectrum overlaid with the 6\arcse $\times $ 6\arcse\, HST/Subaru images. The purple lines show the window of spectra used to extract 1-D spectra. The green lines on the HST(F125)/Subaru(stacked v,b and i band images) images are the LRIS slits on the galaxy. The lower panel shows the extracted 1D spectrum inside the aperture shown with purple lines. The grey region show bootstrapped spectra and the black solid line is the median spectrum of the bootstrapped sample. The red line is the fitted double Gaussian profile.}
\label{fig:field}
\end{center}
\end{figure*}

\subsection{Electron Density}
\label{subsec:elden}
Emission lines originating from collisional excitation and de-excitation are affected by the electron density of the gas cloud. Thus, the electron density of a star-forming galaxy can be estimated using emission line fluxes of two energy levels from the same species that have similar excitation energy but different statistical weight and radiative transition probabilities \citep{1989agna.book.....O}. The emission line flux ratio of the doublet only depends on the electron density and is modelled using collisional strengths and transition probabilities of each component using known atomic data.\\

We use the ratio of emission line doublets \sii\ and \oii\ lines as a function of electron density as derived by \citet[][equation \ref{eq:r_eq}]{Sanders2016}. \cite{Sanders2016} assume a constant temperature of 10,000 K and a typical metallicity of HII regions. The errors in our electron density measurements are larger than the difference introduced by relaxing the constant temperature or metallicity assumption.\\
\begin{equation}\label{eq:r_eq}
R(n_{e}) = a\frac{b+n_e}{c+n_e}
\end{equation}
where, $n_e$ is the electron density of the gas, and $a,b,c$ hold the values listed in table \ref{tab:ratio}.\\

\begin{table}[h]
\begin{center}
\begin{tabular}{lllll}
\hline
$R(n_e)$ &    a      &    b     &    c      \\
\hline \hline
$[OII]$  & 0.3771 & 2,468 & 638.4  \\
$[SII]$  & 0.4315 & 2,107 & 627.1   \\
\hline
\end{tabular}
\end{center}
\label{tab:ratio}
\end{table}

By inverting the above formula, the electron density of the gas can be calculated as:
\begin{equation}\label{eq:ne_eq}
n_e(R) = \frac{cR-ab}{a-R}
\end{equation}

Electron densities derived using eq. \ref{eq:ne_eq} for both \oii\ and \sii\ are similar \citep{Sanders2016}.

To obtain the \oii\ line ratio, we calculate the flux by integrating the fitted Gaussian profile within $3\sigma$ bound for each emission line and calculate the electron density using the equation \ref{eq:ne_eq}. To determine the uncertainty in the electron density, we calculate electron density for each bootstrapped realizations of the observed spectra and take standard deviation of the distribution as $1\sigma$ error on the electron density. For sample sets, we consider the median and error on the median of electron density throughout the paper.

\section{Results and Analysis}

\label{sec:results}
\subsection{Electron Density and Environment}
We measure the electron density for individual galaxies in the $z\sim 1.6$ UDS proto-cluster and field using the ratio of the \oii\,(\lambdaoii) emission lines and equation \ref{eq:ne_eq}. We measure the median electron density for the six cluster galaxies at $z\sim1.62$ of \nel\,=\,\udscls\, \cm\ and for the two field galaxies at similar redshift the average value of \nel\,=\,\udsfl \, \cm\ (Fig.\,\ref{fig:vs_mass}). Although our field value is based on only two galaxies, we stress that the electron density is comparable to that measured by \cite{Kaasinen2016} for field galaxies $z\sim1.5$ (\nel\,=\,$114\pm28$ \cm). Due to limitations in sample size for field galaxies, we combine our field galaxies from LRIS in the UDS field with field galaxies from \cite{Kaasinen2016}. The median electron density for this combined sample is \nel\,=\ \fldmel\ \cm. 

We find tentative evidence of higher electron density in cluster galaxies compared to field galaxies ($\sim 2.6\sigma$). However, we are limited by the  sample size and have significant scatter in the individual electron density measurements to make reliable conclusions (see Fig.\,\ref{fig:vs_mass}b).
We note that our sample is selected to be bright \oii\ emitters, which biases our sample against cluster members that are  undergoing environment dependent evolution and have lower star formation rates. We also note that 2 of the cluster galaxies and both field galaxies are merger components (Fig.\,\ref{fig:cluster},\,\ref{fig:field}). However, we find no significant difference in their electron density compared to the rest of the sample.  

\subsection{Electron Density at $z \sim 0.0$ and $z \sim 1.6$}
For comparison to the local $z\sim0$ sample, we use the \sii\,(\lambdasii) ratio due to the lack of resolved \oii\,(\lambdaoii) doublet in the \sdss. \cite{Sanders2016} show that electron densities measured with \oii\ and \sii\ are consistent, thus are comparable. To measure the redshift evolution of electron densities, we combine the cluster and field samples. The median electron density of the combined $z\sim1.62$ sample is \udsall \,\cm.  Whereas, the median electron density for the local \sdss\ sample is \nel\,=\,$30\pm1$ \cm, showing a nearly 8.5 times increase in the electron density at $z\sim1.5-2$. 

The increase in electron density with redshift when comparing $z\sim0$ sample from \sdss\ to the $z\sim1.5$ sample is significant at $\sim 3.8\sigma$ level. This result is consistent with other studies that also find a high electron density for galaxies in high redshift \citep{Brinchmann2008,Shirazi2013,Sanders2016}. 

The high redshift samples are intrinsically biased towards galaxies with higher SFR compared to the SDSS sample. \cite{Kaasinen2016} find that the rising SFRs with redshift is responsible for the higher electron density of high redshift galaxies. For comparison with the local \sdss\ star-forming galaxies and to correct for the bias of the high redshift galaxies towards higher SFR compared to local \sdss\ galaxies, we select \sdss\ sample in the same SFR range as our $z=1.6$ sample (2.8 \sfr $\leq$ SFR $\leq$ 23.6 \sfr). The median electron density of the SFR matched \sdss\ sample is \nel\,=\,$31\pm 9$ \cm. We find no significant change in the electron density of the local \sdss\ sample even after matching with SFR of our high redshift sample (further discussed in section \ref{subsec:sfr}).

\subsection{Electron Density Vs Stellar Mass}
\begin{figure*}[t]
\centering
\includegraphics[width=\textwidth]{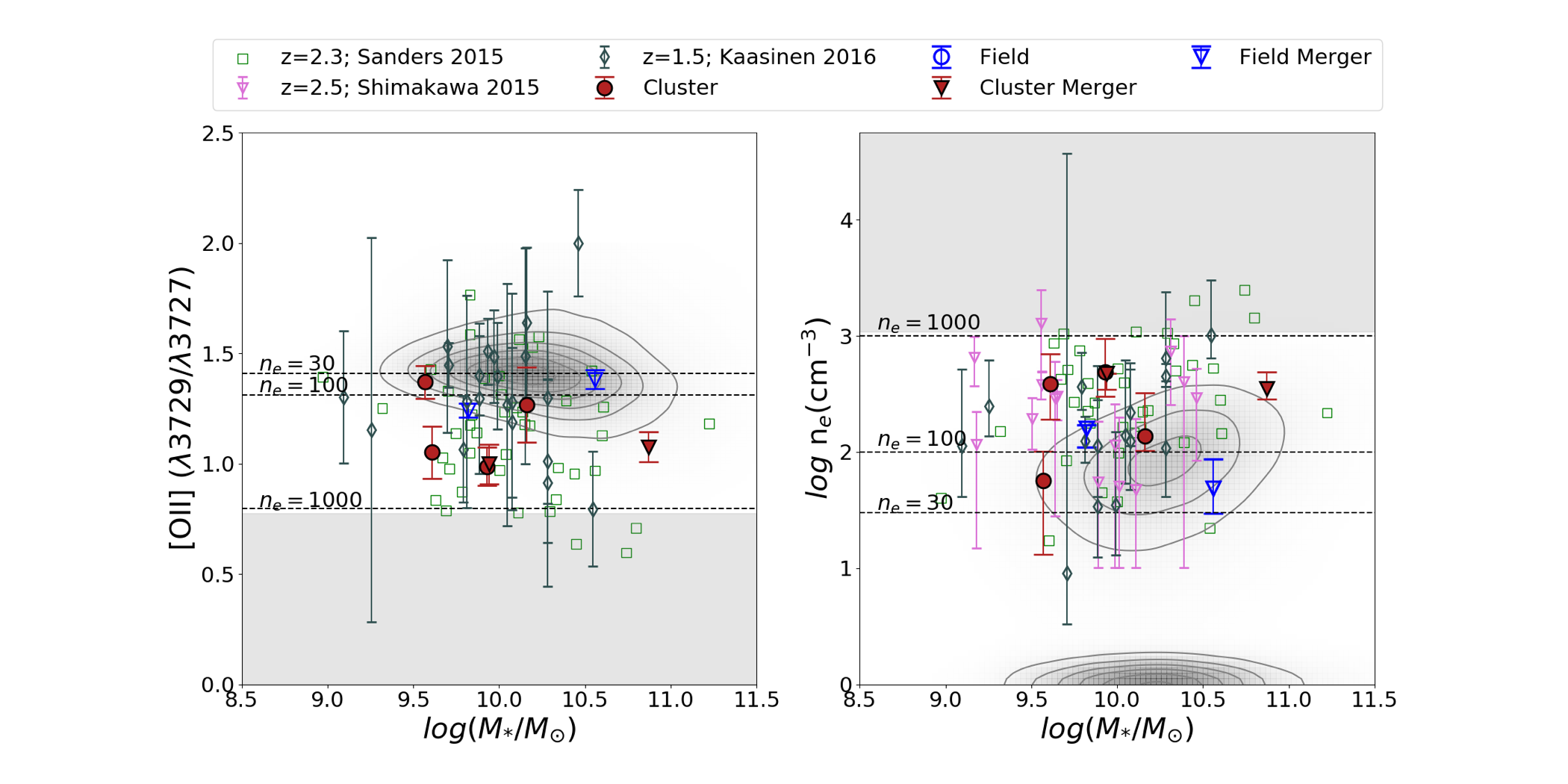}
	\caption{Ratio of \oii\ and \sii\ doublet (a) used to calculate electron density and  Electron density (\nel) (b) as a function of stellar mass (\mstar). Cluster and field galaxies at $z\sim 1.6$ shown by red filled and blue unfilled markers respectively. We compare our results with three different comparison data sets of field galaxies at $z\sim 2.3$, $z\sim2.5$ and $z\sim 1.5$, with green, pink and grey unfilled symbols respectively. The meshed grey contours show the electron density for the \sdss\ sample. Grey shaded area shows the upper-limit of non-detection for the UDS proto-cluster sample.  }
\label{fig:vs_mass}
\end{figure*}

We investigate how the electron density varies with the stellar mass of the galaxy (Fig\,\ref{fig:vs_mass}). The median electron density for the UDS proto-cluster sample at median \mstar\,$=9.93$ with $1\sigma$ scatter of $0.43$ is \nel\,=\,\udscls\,\cm. For our two field galaxies with average \mstar\,$=10.19$ with $1\sigma$ scatter of $0.37$, the average electron density is \nel\,=\,\udsfl\,\cm. At similar stellar mass range, the median electron density of cluster galaxies is at a $\sim 2.6 \sigma$ difference to field galaxies, within the limitation of our sample size.

\begin{figure}[h] 
\centering
\includegraphics[scale=0.35]{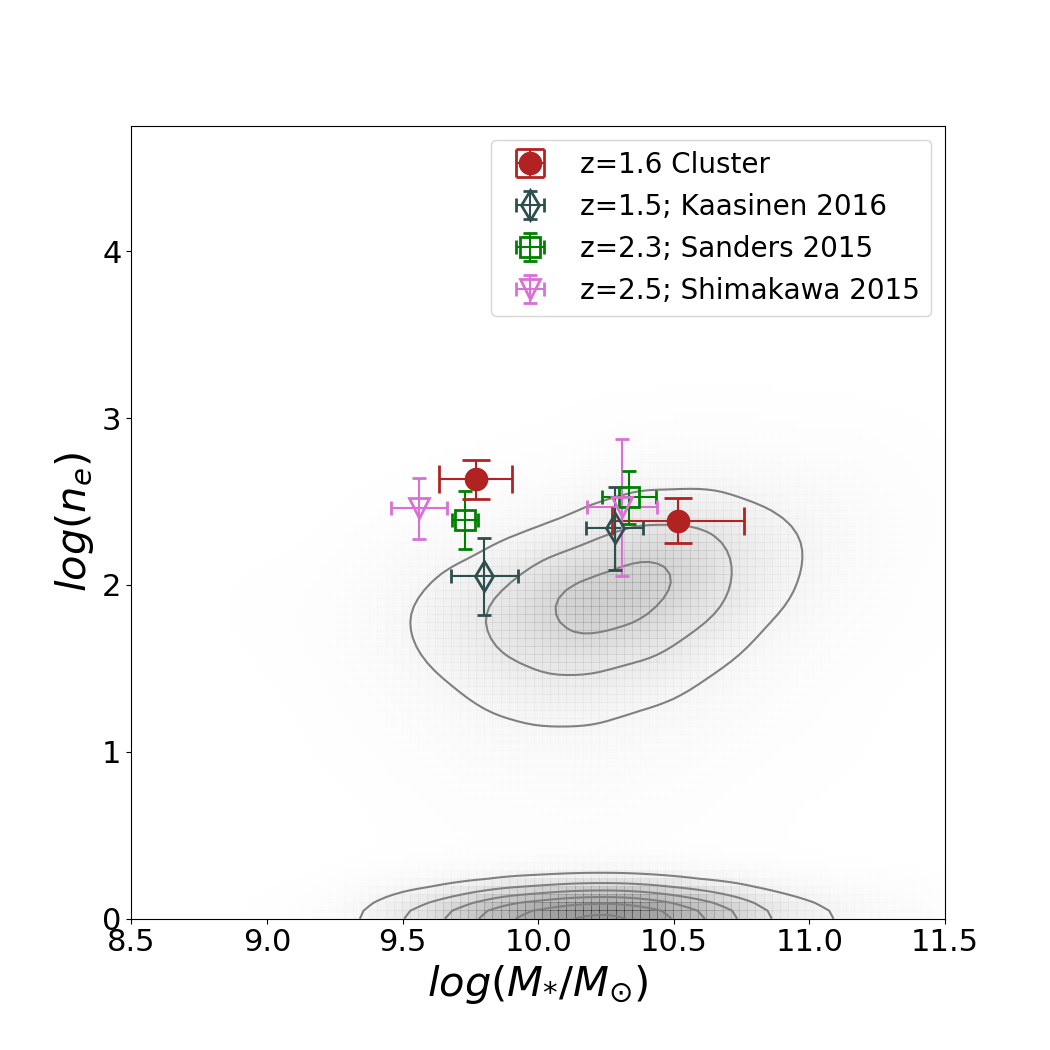}
\caption{Electron density \nel\ as a function of stellar mass (\mstar) for the low mass (\mstar $\leq 10$) and high mass (\mstar$>10$) bins plotted against the median stellar mass of the binned galaxy sample. Cluster galaxies at $z\sim 1.6$ shown by red filled circles. We compare our results with three different comparison data sets of field galaxies at $z\sim 2.3$, $z\sim2.5$ and $z\sim 1.5$, with green, pink and grey unfilled symbols respectively. The meshed grey contours show the electron density distribution for the \sdss\ sample. }These results show no significant variation between electron density of cluster galaxies at $z\sim 1.6$ with high redshift field comparison samples.
\label{fig:vs_mass_bin}
\end{figure}

We bin our sample into two stellar mass bins of \mstar $\leq 10$  and \mstar $>10$ (Fig.\,\ref{fig:vs_mass_bin}). The high mass bin (4 galaxies) of the cluster sample at $z=1.62$ has a median mass of \mstar = 10.5 and median electron density of \nel\,=\,$243 \pm 74 $ \cm\ and the low mass bin (2 galaxies) with median mass of \mstar = 9.77, have median electron density of \nel\,=\,$429 \pm 116 $ \cm. Due to the limited number of field galaxies in our sample, we compare our results with the field galaxy sample from \cite{Kaasinen2016} at $z=1.5$. The median stellar mass of the high stellar mass bin in \cite{Kaasinen2016} is \mstar = 10.28 and electron density is \nel\,=\,$218 \pm 19\,$\cm. Similarly, the low stellar mass bin in \cite{Kaasinen2016} has a median mass of \mstar = 9.8 and median electron density of \nel\,=\,$113 \pm 46$ \cm.

We find no significant correlation ($<2\sigma$) between the electron density and stellar mass. Although, we see a reversal in trend between cluster galaxies at $z=1.6$ and comparison field sample at $z=1.5$ \citep{Kaasinen2016}, the differences are within $2 \sigma$ level and hence not statistically significant. Our result is consistent with other high redshift observations \citep{Kaasinen2016,Sanders2016,Shimakawa2015}. 

\subsection{Electron Density Vs Star Formation Rate}
 \label{subsec:sfr}
\begin{figure*}[t] \centering
\includegraphics[width=\textwidth]{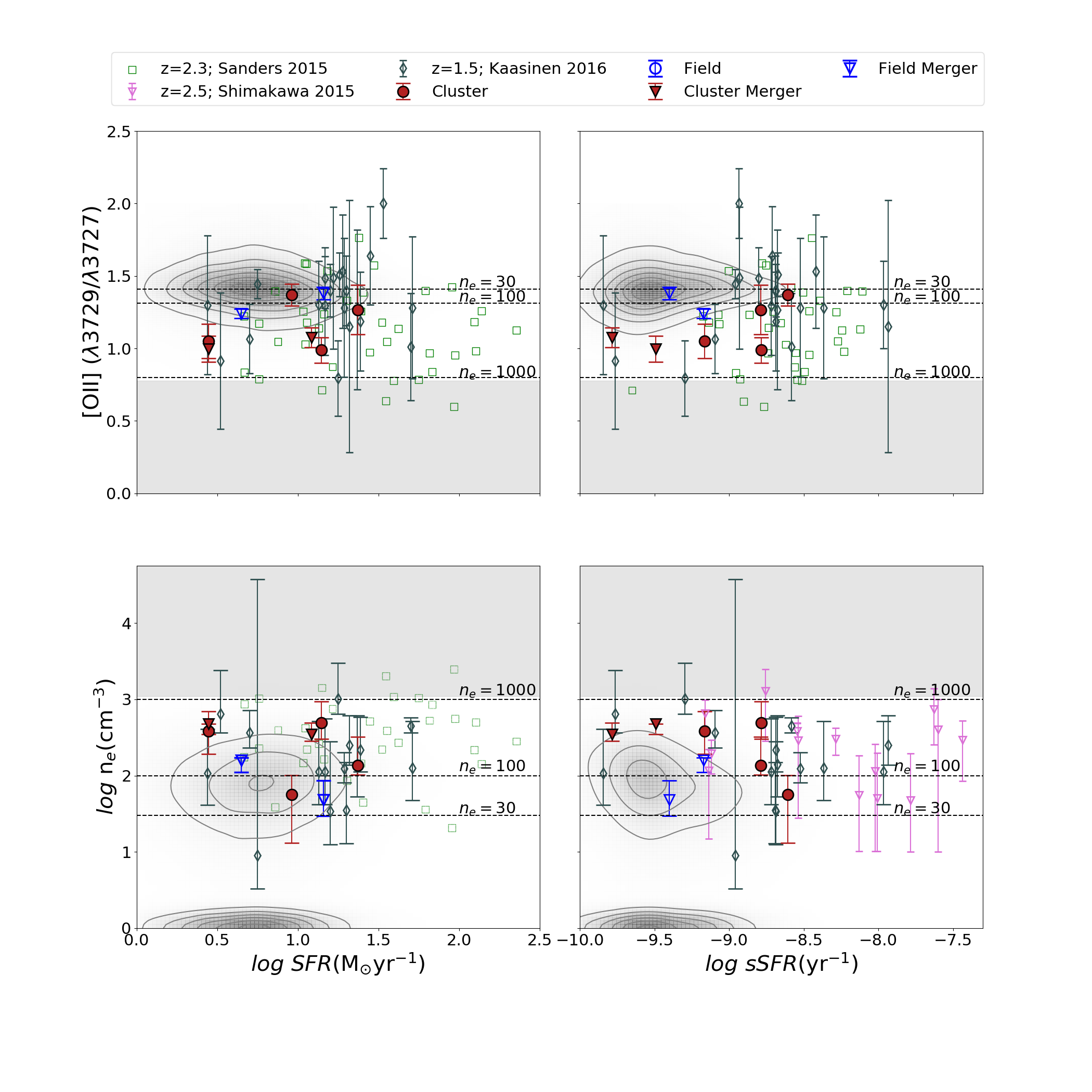}
	\caption{Ratio of \oii\ or \sii\ doublet (upper panels) and Electron density (\nel) (lower panels) as a function of log star formation rate (\sfr) (left) and specific star formation rate ($yr^{\rm{-1}}$) (right). Cluster and field galaxies at $z\sim 1.6$ shown by red filled and blue unfilled markers respectively. We compare our results with three different comparison data sets of field galaxies at $z\sim 2.3$, $z\sim2.5$ and $z\sim 1.5$, with green, pink and grey unfilled symbols respectively. The meshed grey contours show the electron density distribution for the \sdss\ sample. We measure no correlation of electron density with the SFR or sSFR and find no significant variation between electron density of cluster galaxies at $z\sim 1.6$ with high redshift field samples.}

\label{fig:vs_sfr}

\end{figure*}

We analyze the correlation of electron density with the star formation rate (SFR) and specific star formation rate (sSFR) in Fig.\,\ref{fig:vs_sfr}. At $z=1.6$, the cluster and field sample have a median SFR of 10.6 \sfr\ with $1\sigma$ spread of 7.8 \sfr\  and 9.4 \sfr\ with $1\sigma$ spread of 7.1 \sfr\ respectively. We continue to find tentative dependence of electron density on environment in cluster and field sample at $z\sim 1.6$, however, we are limited by large associated errors and small sample size. We also find no significant correlation between the SFR and electron density in our $z=1.6$ sample, consistent with results from \cite{Sanders2016} and \cite{Kewley2013}. \cite{Shimakawa2015} find a positive correlation between the electron density and sSFR at a 4$\sigma$ level, albeit with large error bars and limited sample at $z\sim 2.5$.

For comparison with the local \sdss\ star-forming galaxies and to correct for the bias of the high redshift galaxies towards higher SFR compared to local \sdss\ galaxies, we select \sdss\ sample in the same SFR range as our $z=1.6$ sample (2.8 \sfr $\leq$ SFR $\leq$ 23.6 \sfr). The median electron density of the SFR matched \sdss\ sample is \nel\,=\,$31\pm 9$ \cm. The electron density of SFR matched \sdss\ sample by \cite{Kaasinen2016} is \nel\,=\,$98 \pm 4$ \cm, similar to the electron density of $z\sim1.5$ sample in their study. \cite{Kaasinen2016}  selected the SFR matched \sdss\ sample by matching the distribution in the SFR between $z\sim1.5$ and the local sample. However, our limited sample at $z\sim 1.6$ does not allow us to match the distribution of SFRs. The different SFR distribution between the SFR matched \sdss\ sample and  our $z\sim 1.6$ sample can contribute to the observed difference in their median electron density.

\section{Discussion}
\label{sec:disc}

We measure the electron density for six galaxies in the UDS proto-cluster at $z\sim 1.6$ and compare  it with field galaxies at $z\sim 1.5$. We find that cluster galaxies have higher electron density compared to field galaxies ($\sigma \sim 2.6$). However, the small sample size and large scatter in individual electron densities make our conclusions tentative only. Our results are different to \cite{Kewley2015}, who do not find significant effect of environment on electron density in the COSMOS proto-cluster at $z\sim2.0$. We note the difference in methods for calculating electron densities by \cite{Kewley2015}, who use \sii\ emission lines and stacking of 1D spectra to increase the SNR.

In contrast to our results, by stacking galaxies in stellar mass, SFR and sSFR bins \cite{Darvish2015} measure $\approx 17$ times lower electron density for galaxies in a filamentary region ($\approx 5$ times denser than the field) compared to field galaxies at $z\sim 0.5$. However, their individual electron density measurements have significantly large errors and scatter. Moreover, we are looking at environmental dependence on the electron density at $z\sim1.5$ where environmental effects are less significant as opposed to $z\sim 0.5$ \citep{Kewley2013,Tran2015, Gupta2018, Alcorn2019}. 

We observe redshift evolution of the electron density between the local \sdss\ sample and $z\sim 1.6$ sample after matching the stellar mass, SFR and sSFR range of the two samples. We find that electron density increases by a factor of $\approx 8.5$ from $z\sim0$ to $\sim1.5$, even with our limited sample size. \cite{Kaasinen2016} find that after matching the SFR distribution between the local \sdss\ galaxies with galaxies at $z\sim 1.5$, difference between the electron density of low and high-redshift sample disappears. Different methods for selecting SFR-matched sample from local and $z\sim 1.6$ sample might be responsible for this observed difference (Section \ref{subsec:sfr}). 

Our work indicates no apparent correlation between the electron density and the stellar mass, SFR or specific SFR of galaxies at $z=1.62$.  Cluster galaxies in the low stellar mass bin have slightly higher in electron density compared to field galaxies, however the difference is at $<2\sigma$ significance (Fig.\,\ref{fig:vs_mass_bin}).   The higher SFR and gas surface density of galaxies at $z\sim 1.6$ compared to galaxies in the local universe  might be responsible for $\approx 8.5$ times increase in the  electron density of galaxies at $z\sim 1.6$ \citep{Madau2014}.

By analyzing the Subaru and HST imaging, we find that both field galaxies and two of six cluster galaxies (Fig.\,\ref{fig:cluster},\,\ref{fig:field}) in our sample are part of merger pairs. Within the small sample, the electron density of mergers are comparable to the rest of the sample at  $z\sim 1.5$. Merging galaxies have SFRs comparable to the non-merger sample, which might be responsible for their similar electron densities. Mergers have $\approx 1$\,dex lower sSFR than the rest because of their higher stellar masses.
However, we require larger sample of mergers to fully investigate the role of mergers on the electron density of galaxies.

Electron density measured using different species ratios probe different parts of the HII regions in the galaxy. In a recent paper \cite{Kewley2019b} find that electron densities measured using \sii\ ratios would probe the outer parts of the nebulae in the high pressure clumps unlike \oii\ ratio. However, \cite{Sanders2016} find no significant difference between electron densities calculated using \oii\ and \sii.

Studies like ours that measures the electron density in intermediate and high redshift universe remain challenging. The large sample of proto-clusters  at $z>1.0$ needs to be explored to fully understand the role of environment on the electron density. Also, we currently do not understand how diffused-ionised gas emission effects the electron density measurements from the integrated emission line studies \citep{Shapley2019}. Near-infrared spectrographs on next generation space and ground based telescopes would be able to provide sub-kpc scale resolution on intermediate and high-z galaxies to further analyze the redshift and environment dependent evolution of the electron density galaxies.

\section{Summary}
\label{sec:sum}

We analyze how environment affects the electron density of galaxies in the UDS proto-cluster (IRC0218) at $z=1.6$.
We use spectroscopic data from LRIS on Keck1 taken as part of the \zfire\ survey and calculate the electron density using the ratio of optical emission lines \oii\,(\lambdaoii). We identify six cluster members ($1.6118 <z_{spec}< 1.6348$) and two field galaxies with resolved \oii. We compare our results with the \sdss\ DR7 emission line catalogue from the local universe, and other field samples at $z\sim 1.5 - 2.5$ from literature. We note that our $z=1.6$ sample is biased towards galaxies with higher SFR compared to local \sdss\ sample.

With our limited sample at $z=1.62$, we measure the median electron density of the cluster galaxies to be \udscls\, \cm\, and \udsfl\,\cm\, for the field sample. Despite the higher electron density measured in the cluster, the difference is statistically insignificant due to high associated errors and limited sample size. We find a large scatter in the electron density of galaxies, similar to the local \sdss\ and other $z>1.5$  samples.

We find that the average electron density increases with increasing redshift. The median electron density in local \sdss\ star-forming galaxies is measured as $30 \pm 1\ \rm{cm}^{-3}$ and the median electron density of $z=1.62$  sample is \udsall\,\cm. We also find no significant correlation between the electron density  and stellar mass (Fig.\,\ref{fig:vs_mass} and Fig.\,\ref{fig:vs_mass_bin}), SFR  and sSFR (Fig.\,\ref{fig:vs_sfr}), in agreement to other studies at $z>1.5$.

To summarize, we find tentative evidence of effect of environment on the electron density of galaxies at $z=1.62$. However we note that we are limited by a small sample size of eight galaxies. Further investigation of electron density with a larger sample for clusters at $z>1.0$ and higher SNR spectra are needed to establish conclusively any possible effect of environment on the electron density. 
\newline

K. Tran acknowledges support by the National Science Foundation under Grant Number 1410728.
T.Y. acknowledges support from an ASTRO 3D fellowship. GGK acknowledges the support of the Australian Research Council through the Discovery Project DP170103470. T.N. acknowledges the Nederlandse Organisatie voor Wetenschappelijk Onderzoek (NWO) top grant TOP1.16.057. The authors wish to recognize and acknowledge the very significant cultural role and reverence that the summit of Mauna Kea has always had within the indigenous Hawaiian community. 
We are most fortunate to have the opportunity to conduct observations from the summit.

\begin{table*}[t]
\small
	\begin{center}
		\caption{Cluster Galaxies}
		\label{tb:cluster galaxies}

        \begin{tabular}{ p{1.2cm}  p{1.5cm}  p{1.7cm} p{1.2cm}  p{2.1cm} p{1.6cm} p{1.8cm} p{2.2cm} p{1.2cm}  }

        \hline
       Obj ID & RA$^a$      & DEC $^b$     & ${\rm z}_{spec}$\ $^c$ & \mstar  $^d$ & log SFR $^e$ & log sSFR $^f$ & Ratio $^g$ &${\rm n}_e$\ $^h$   \\ \hline
        39463 & 2:18:22.3 & -5:10:34.5 & 1.6220      & 9.57                          &0.96 &-8.60& $1.429 \pm 0.091$ &$57^{+32}_{-83}$                                       \\ 
        40243   & 2:18:28.0 & -5:10:10.5 & 1.6220      & 9.61                          &0.45 &-9.16& $1.068 \pm 0.199$ &$384^{+232}_{-266}$                      \\
        41297   & 2:18:24.2  & -5:09:39.5 & 1.6221     & 9.93                         & 1.15& -8.78& $0.985 \pm 0.148$ & $491^{+319}_{-241}$                      \\
        47191   & 2:18:29.8 & -5:06:38.5 & 1.6331     & 9.94                         & 0.45 &-9.49& $0.958 \pm 0.056$ &$474^{+0.47}_{-147}$                       \\ 
        46922   & 2:18:26.8 & -5:06:49.4 & 1.6302     & 10.16                        & 1.37 &-8.79& $1.284 \pm 0.080$  &$137^{+117}_{-39}$ \\
        38455   & 2:18:26.2 & -5:11:10.5 & 1.6238     & 10.87                         &1.09&-9.78& $1.086 \pm 0.080$ &$349^{+119}_{-71}$ \\ \hline

        \end{tabular}
    \end{center}
\end{table*}

\begin{table*}[t]
\small
	\begin{center}
		\caption{Field Galaxies}
		\label{tb:field galaxies}

        \begin{tabular}{ p{1.2cm}  p{1.5cm}  p{1.7cm} p{1.2cm}  p{2.1cm} p{1.6cm} p{1.8cm} p{2.2cm} p{1.2cm}  }

        \hline
        Obj ID & RA$^a$      & DEC $^b$     & ${\rm z}_{spec}$\ $^c$ & \mstar  $^d$ & log SFR $^e$ & log sSFR $^f$ & Ratio $^g$ &${\rm n}_e$\ $^h$   \\ \hline
        44518 & 2:18:22.3 & -5:10:34.5 & 1.4950      & 10.56                          &1.16 &-9.40& $1.381 \pm 0.036$ &$49^{+28}_{-24}$                                       \\ 
        49505   & 2:18:28.0 & -5:10:10.5 & 1.4068  & 9.82                          & 0.65 &-9.17& $1.257 \pm 0.044$ &$160^{+12}_{-59}$\\ \hline

        \end{tabular}
    \end{center}
	
	\begin{flushleft}
		{\bf Notes:}\\
		$^{\rm a}$ Right accession (J2000)\\
		$^{\rm b}$ Declination (J2000)\\
		$^{\rm c}$ spectroscopic redshift\\
		$^{\rm d}$ log stellar mass (from \candels survey)\\
		$^{\rm e}$ log star formation rate (from \candels survey) [\sfr]\\
		$^{\rm f}$ log specific Star formation rate (SFR/stellar mass) [${\rm yr}^{-1}$]\\
		$^{\rm g}$ Ratio of \oii \ emission lines ($\lambda3729/\lambda3726$)\\
		$^{\rm h}$ Electron Density $(\rm{cm}^{-3})$ calculated from ratio of [OII] doublet with $1\sigma$ errors\\
		 
	\end{flushleft}

\end{table*} 

%results table
\begin{table*}[t]
\small
	\begin{center}
		\caption{Median Electron Density Measurements}
		\label{tb:elden}

        \begin{tabular}{lll}
         	\hline
	        Sample Set & redshift z & \nel (\cm) \\ \hline
	        UDS proto-cluster &1.62& \udscls  \\
	        Field sample &$\sim 1.5$ & \udsfl \\
	        Field sample + \cite{Kaasinen2016} &$\sim 1.5$ & \fldmel \\
	        Full sample &$ \sim1.5 $& \udsall \\
	        \sdss & $<0.1$ & $30\pm1$ \\ 
	        \cite{Kaasinen2016} & $1.5 $& $114\pm28$ \\
	        \cite{Sanders2016} (\oii) & $2.3$ & $225^{+119}_{-4}$ \\ \hline

        \end{tabular}
    \end{center}

\end{table*} 

\bibliographystyle{aasjournal}

\end{document}